# Predicted high-temperature superconductivity in rare earth hydride ErH$_2$ at moderate pressure

Yiding Liu,[1,2] Qiang Fan,[3] Jianhui Yang,[1] Lili Wang,[4] Weibin Zhang,[5] Gang Yao,[6,*]

[1] College of Mathematics and Physics, Leshan Normal University, Leshan 614004, China.
[2] Institute of Atomic and Molecular Physics, Sichuan University, Chengdu 610065, China.
[3] School of New Energy Materials and Chemistry, Leshan Normal University, Leshan 614004, China.
[4] Institute of Computer Application, China Academy of Engineering Physics, Mianyang 621900, China.
[5] College of Physics and Electronics Information, Yunnan Key Laboratory of Optoelectronic Information Technology, Yunnan Normal University, Kunming, 650500, China.
[6] Tsung-Dao Lee Institute, Shanghai Jiao Tong University, Shanghai 200240, China.

**Abstract:** Hydrides offer an opportunity to study high-temperature ($T_c$) superconductivity at experimentally achievable pressures. However, they remained extremely high. Using density functional theory calculations, herein we demonstrated that a newly rare earth hydride, namely bulk ErH$_2$, could be superconducting with a $T_c$ around 80 K at 14.5 GPa. To date, the drived pressure is the lowest reported value for compressed hydrides. Besides superconductivity, Fermi Surface nesting and Kondo effect were manifested at this pressure. Intriguingly, due to Kondo destruction, superconductivity was prone to exist at 15 GPa. Under the rest of applied pressures, we also revealed a gap of band structure at 20 GPa on the background of normal metallic states. At 20 GPa, this compressed system could act as a host of superconductor being judged from a sharp jump of spontaneous magnetic susceptibility with an evanescent spin density of state at Fermi level along with the competition between spin density wave and superconductivity. Finally, electron pairing glue for ErH$_2$ at these three typical pressures was attributed to the antiferromagnetic spin fluctuation.

## 1. Introduction

Hydrogen in certain hydrides can become metallic at a lower pressure than pure hydrogen, presumably because of additive effects from chemical precompression.[1] These hydrides might well be superconducting up to high critical transition temperature (high-$T_c$). Several theoretical and experimental efforts were underway to better understand the high-$T_c$ superconductivity of such hydrides. These compounds include $GeH_4$, (64 K at 220 GPa),[2] $YH_3$, (40 K at 17.7 GPa),[3] $H_3S$, (203 K at 90 GPa),[4] $VH_8$, (71.4 K at 200 GPa),[5] $LaH_{10}$, (260 K at 180-200 GPa,[6] and 250 K at 170 GPa[7]), and other room-temperature superconductors $YH_{10}$, (303 K at 400 GPa),[8] $Li_2MgH_{16}$, (473 K at 250 GPa).[9] Note that in all the hydrides mentioned above, the major glue for electron pairing is phonon.

Among rare-earth metals, Er showed a powerful getter for hydrogen.[10] The stoichiometry erbium dihydride ($ErH_2$) and its nonstoichiometric composition $ErH_{2+x}$ (-0.15< $x$ < 0.15) both crystallize with fcc-$CaF_2$ type crystal structure (space group: *Fm-3m*), which has often been described as the *β*-Phase.[11,12] The conventional unit cell of $ErH_2$ is shown in Fig. 1 in which the H atoms occupied the tetrahedral sites of the Er fcc lattice. In the meantime, the *f* electrons in lattice are often neither fully localized around their host nuclei nor fully itinerant. This localized versus itinerant duality has proposed the involvement of *f* electrons as valence ones in the systems, that is, strongly correlated-electron materials, therefore promoting the emerging researches of heavy fermion superconductors.[13] On the other hand, it was argued that antiferromagnetic (AFM) spin fluctuation was another plausible pairing mechanism driven by Coulomb repulsion in this system, and could lead to high-$T_c$ superconductivity.[14-17] The Hubbard model could be consequently applied to this system.[18-20] AFM spin fluctuation could be reflected with mutation from a large Fermi Surface (FS) to a small one,[21] and a jump of spin DOS at Fermi level ($E_F$) or spontaneous magnetic susceptibility ($\chi_s$).[15,17] It is worth investigating whether a compressed system can be superconducting and if there is another pairing mechanism. Despite the investigation discussed above for *f* electrons in a crystal, a systematic study of possible pressure-induced superconductivity in $ErH_2$ is still missing.

In this work, we use first-principles approaches to investigate the high-$T_c$ superconductivity of $ErH_2$, a heavy fermion hydride under pressures. Our purposes were to predict a relatively high $T_c$ at moderate pressure, and to find several accompanying physical behaviors, such as FS nesting, Kondo effect, Kondo destruction and a gap of band structure on the background of normal metallic states under several corresponding typical pressures.

## 2. Methods

Based on Density Functional Theory (DFT), the present calculations were performed using the Cambridge Serial Total Energy Package (CASTEP) code.[22] The exchange and correlation interactions were described by the Generalized Gradient Approximation (GGA) of Perdew and Wang (PW91).[23] Norm-conserving pseudopotentials of Er and H were taken from the CASTEP library to describe the properties of the perfect crystal. The valence electron configurations of Er and H were $4f^{11}5d^{1}5s^{2}5p^{6}6s^{2}$ and $1s^{1}$, respectively. The geometry optimizations of the shape and size of its unit cell were performed using the plane-wave basis sets with energy cutoff of 380 eV, and a 10 × 10 × 10 Monkhorst–Pack (MP) grid sampling the Brillouin Zone (BZ), i.e. k-points which was 36 × 36 × 36 for electronic structure calculations including FSs and Electron Localization Functions (ELFs). Spin polarization was taken into account for all valence electrons. The Hubbard $U$ model for strongly correlated Er-$f$ electrons was considered, and on-site $U$ = 6 eV. The phonon calculations were performed using finite displacements method[24] with the same precision settings of its geometry optimizations. All above parameter settings and their values were verifiable to convergence tests.

## 3. Results and discussion

ErH$_2$ was experimentally confirmed to possess AFM[25] that was subjected to have an intimate link with both FS nesting[26] and superconductivity.[15] The FS topologies were calculated out in this work accompanying with their electronic Density of States (DOS) and band structures under ambient pressures ranging from 0 to 21.5 GPa.

The band structures under 0, 14.5, 15 GPa and other pressures except 20 GPa were respectively shown in Fig. 2a-d. The results of FSs appeared three types of topology as shown in Fig. 3a-d, electronlike-type under 0 GPa, holelike-type under 14.5 and 15 GPa (large and small respectively) and open-type at other pressures except for 20 GPa under which DOS and band structure revealed a gap, thus showed no FS around its $E_F$. Compared with 0 GPa, at which the corrugated FS reflected the influence of phonon on the properties of electrons around $E_F$, the other smooth FSs in Fig. 3 indicated that the interaction between phonon and electrons decreased considerably.[27]

The topology of the holelike FS at 14.5 GPa (Fig. 3b) did show nesting property which was in accordance with its AFM.[25] This consistency between FS nesting and AFM had also been confirmed in the conclusion for cubic structural ErGa$_3$.[26] This FS was apparently the largest one in the whole applied pressures (Fig. 3a-d). Moreover, in the conduction band region at this pressure (Fig. 4c), there was a sharp spin-down $f$ ($\beta$) peak between 4-5 eV just corresponding to the spin-up conduction $d$ ($\alpha$) electrons in this energy scale. Hence, the state at 14.5 GPa (Fig. 3b) could infer the Kondo effect,[13] in which the fully itinerant $f$ ($\beta$) electrons like magnetons immersed in conduction $d(\alpha)$ electrons, entangled with them sufficiently, and contributed to the Fermi volume.[13,28] This itinerancy of $f$ electrons was also embodied with the localized $s$ electrons of H in the ELF at this pressure (Fig. 4a). Compared with the ELFs under other pressures (Fig. 4b) where electrons localized around Er, the ELF at 14.5 GPa (Fig. 4a) showed a distinct localization of electrons around H. Judging from Fig. 4c in which two large H-derived sharp $s$ electrons DOS peaks (spin-up ($\alpha$) and spin-down ($\beta$) respectively) were most adjacent to the $E_F$, H were essentially isolated like atomic

H in this lattice. Accordingly, the full formation of a large FS could correspond to heavy fermion superconductivity.[21] Meanwhile, Fig. 4e demonstrated the variation of spin DOS at $E_F$ and the $\chi_s$ along with the applied pressures. Combining with the explicit hopping of spin DOS at $E_F$ at 14.5 GPa without abrupt jump of $\chi_s$ (Fig. 4e) and the largest FS with its nesting (Fig. 3b), superconductivity could be implied at 14.5 GPa being mediated by AFM spin fluctuation.[14-17]

Fermi Liquid (FL) could easily exist in this state.[13] Furthermore, basing on phonon calculation, the lattice specific heat at 14.5 GPa was also calculated out.[29] The specific heat ($C$) in the form of $C/T$ dependence on $T^2$ being shown in Fig. 5 under ambient pressures, was obtained with formula (1):[30,31]

$$C = \gamma_1 T^3 + \gamma_2 T \qquad (1)$$

Here, the first term is lattice specific heat, the second term is electrons contribution. $\gamma_1$ is interatomic force constant, $\gamma_2 = \frac{\pi^2}{3} k_B^2 g(E_F)$, $g(E_F)$ is the DOS at $E_F$, $k_B$ is Boltzmann constant. For FL state, the electrons specific heat depends on $T$ linearly, i.e. $\gamma_2 T$. With these regards, high-$T_c$ superconductivity could be predicted with $T_c$ = 78.9 K from the kink point of the curve at 14.5 GPa.[27] This $T_c$ value was around boiling point of liquid nitrogen.

In contrast to Kondo effect, superconductivity could also be driven by Kondo destruction transforming from a large FS to a small one in heavy fermion superconductors, such as $CeCu_2Si_2$.[21] The FS of $ErH_2$ at 15 GPa collapsed abruptly as shown in Fig. 3c which was identified as destruction of Kondo effect.[13,21,32] Hence, this quantum criticality at 15 GPa could also be prone to possess superconductivity. This phenomenon implied the fully localization of $f$ electrons around their host nuclei, i.e. Er, as the ELF (Fig. 4b) had shown. The fact that the localized $f$ electrons do not contribute to the Fermi volume[13,28] was consistent with the small FS (Fig. 3c).

Subsequently, with the Kondo destruction (Fig. 3c), the Ruderman-Kittel-Kasuya-Yosida (RKKY) interaction played a dominant role that led to the crossover feature from FL at 14.5 GPa to Non-Fermi-Liquid (NFL) behavior at 15 GPa where the indirect interaction among the localized $f$ electrons with magnetic moments was mediated mainly by the conduction $d$ electrons.[13,21] The DOS was shown in Fig. 4d. Therefore, despite the superconductivity as it was predicted at 15 GPa, the dependence of electrons specific heat on $T$ was nolinear,[30,31] so the $T_c$ value at 15 GPa could not be obtained from the kink point of the $C/T$ versus $T^2$ curve in Fig. 5. This superconductivity could also be inferred with the medium of AFM spin fluctuation from the noticeable hopping of spin DOS at $E_F$ between 14.5 and 15 GPa without steep variation of $\chi_s$ in this pressure region (Fig. 4e) as well as this FS jump.[14-17] The accompanying behaviors which could also to some extent support the affirmation of superconductivity at 15 GPa were presented in the Supplemental Material.[33]

As it can be prominently seen in DOS and band structures of this bulk (Fig. 6a-c), there was a gap of 6.6 eV at 20 GPa in the present calculations on the background of normal metallic states under the rest of applied pressures. It could be seen in Fig. 6b that the itinerant electrons were from *s* electrons of H and *s*, *p*, *d* ones of Er, whose all spin with up and down direction had almost the same weight and symmetric distribution. The fully localized electrons were from *f* electrons of Er as shown in the ELF (Fig. 4b) and the DOS (Fig. 6b). This manifested that itinerant AFM at 20 GPa vanished abruptly, leading to a localized AFM insulating state.[34] This transition resulted in a sharp jump of $\chi_s$ and an evanescent spin DOS at $E_F$ (Fig. 4e) exclusively at this pressure which could also signify an AFM spin fluctuation. This bulk under 20 GPa could thus act as a host of superconductor owing to this pairing glue.[14-17] This implication was supported by the competition between Spin Density Wave (SDW) and superconductivity.[13,35-37] The existence of SDW could be determined with Pauli paramagnetic susceptibility ($\chi_P(0)$) which was expressed as follow:[34]

$$\chi_P(0) = 2\mu_B^2 \mu_0 g(E_F) \qquad (2)$$

Where, $\mu_B$ is Bohr magneton, $\mu_0$ is vacuum permeability, $g(E_F)$ is the DOS at $E_F$. The SDW was suppressed from the zero DOS at $E_F$ (Fig. 6b) and the formula (2). This hypothesis could also be supported by the tiny and clear deviation of the $V/V_0$ versus pressures relation calculated directly with DFT at this pressure away from the monotonous trend of the curve fitted with *E-V* data of this work (see Fig. S1 in the Supplemental Material[33]). The DOS (Fig. 6b) displayed that the hybridization between localized *f* moments and itinerant electrons is very weak. This could infer the NFL behavior at this pressure.[13,36] Meanwhile, as shown in Fig. 6a and c, the band structures of (110) surface of this bulk at 20 GPa with the range from -6 to 6 eV exhibiting metallic, came very within the gap of this bulk. This metallic surface state with bulk insulating state emerged also in the archetype Kondo insulator, i.e. $SmB_6$.[32] Our results could postulate a possibility for topological superconductor which could host Majorana Zero Mode.[38,39]

### 4. Conclusion

In summary, we revealed that superconductivity emerged in $ErH_2$ under moderate pressures which were 14.5, 15 and 20 GPa with DFT calculations. We proposed that $ErH_2$ at 14.5 GPa was a potential high-$T_c$ superconductor whose $T_c$ was around 80 K. Our current result of the gapped bulk state with metallic surface state at 20 GPa could be conducive to the exploration of topological superconductor which will stimulate further experimental and theoretical studies for the feasibility of $ErH_2$ to host Majorana Zero Mode.

**Corresponding Author**
[*]E-mail: yaogang1257@sjtu.edu.cn

**ORCID**
Yiding Liu: 0000-0003-2265-1827
Gang Yao: 0000-0002-4944-4118

**Notes**
The authors declare no competing financial interest.

**Acknowledgements**
This work was supported by the National Natural Science Foundation of China (Grant No. 12104294). This work was carried out at Shanxi Supercomputing Center of China, and the calculations were performed on TianHe-2.

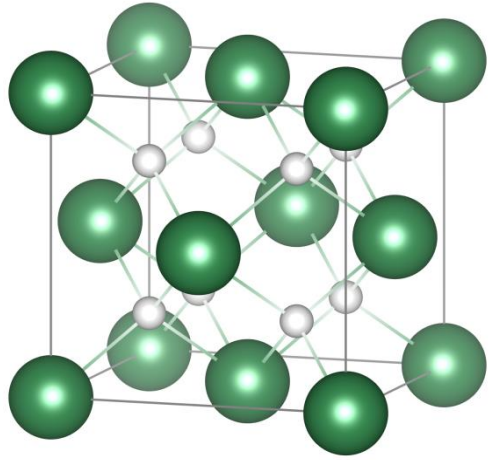

**Figure 1.** A unit cell of ErH$_2$ crystal was displayed. The larger green atoms represented the erbium fcc lattice. The eight smaller gray atoms represented hydrogen occupying all of the tetrahedral sites.

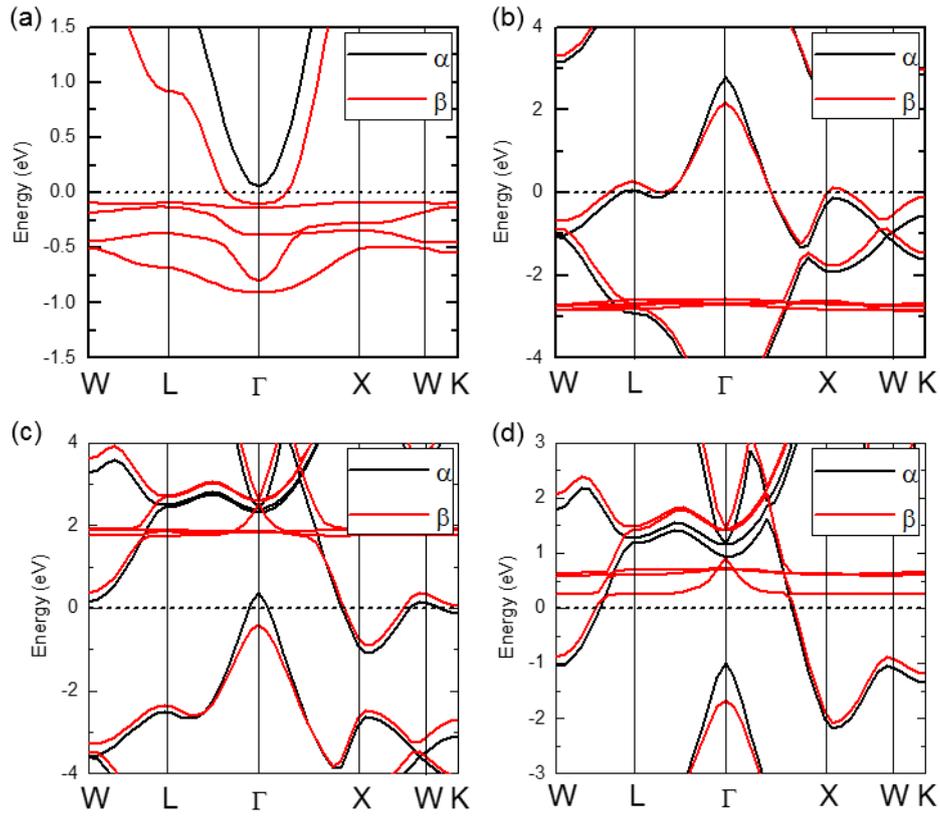

**Figure 2.** Band structures of ErH$_2$ under ambient pressures ranging from 0 to 21.5 GPa. (a) 0 GPa, (b) 14.5 GPa, (c) 15 GPa, (d) 0.5 GPa. At the rest of this pressure range except 20 GPa, band structures were similar with (d), $\alpha$ and $\beta$ denoted spin up and down respectively.

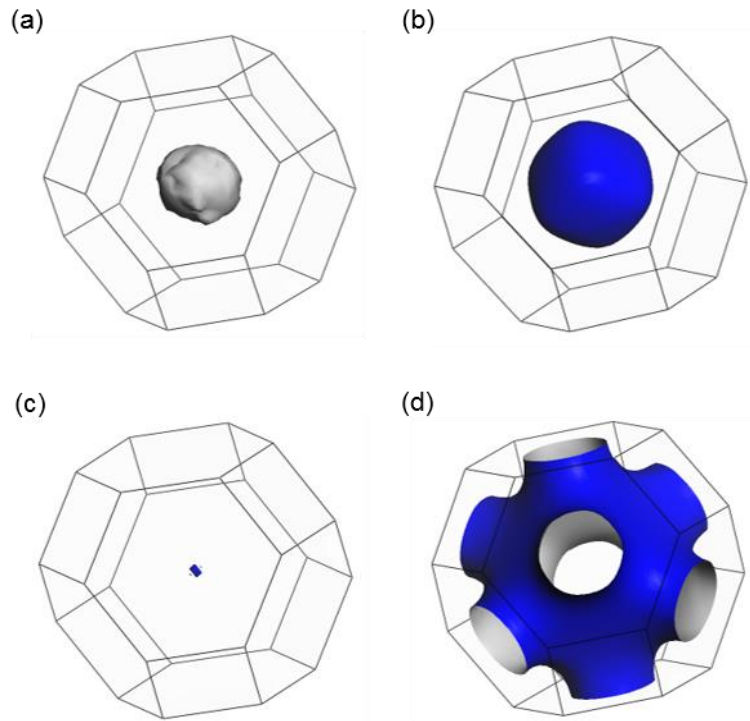

**Figure 3.** Types of FSs for ErH$_2$ at ambient pressures ranging from 0 to 21.5 GPa. (a) electronlike-type under 0 GPa. (b), (c) holelike-type at 14.5 and 15 GPa respectively. (d) open-type at 0.5 GPa. At the rest of this pressure range except 20 GPa, the FSs were open-type similar with the one at 0.5 GPa.

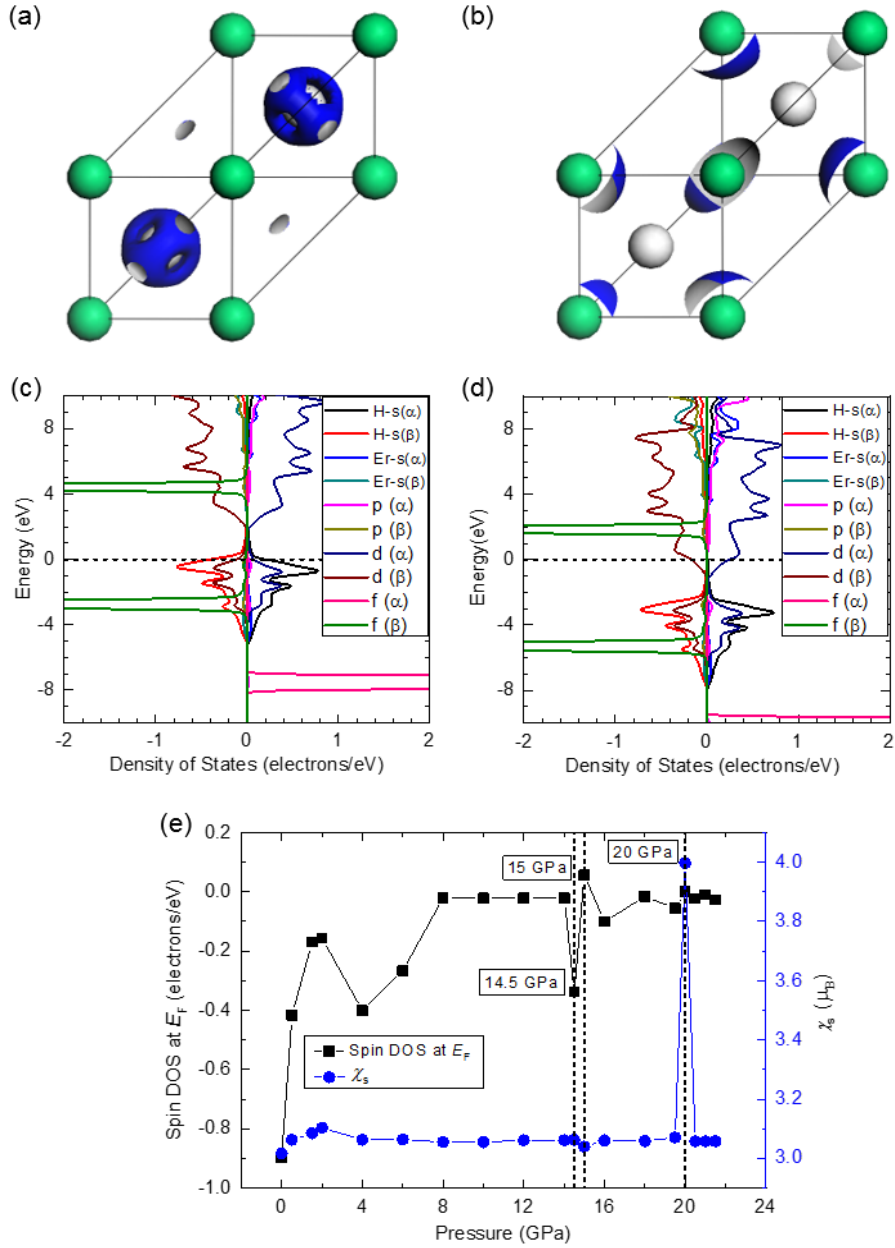

**Figure 4.** (a) Schematic diagram of Electron Localization Function (ELF) at 14.5 GPa. (b) The ELFs under other pressures of the whole applied pressures. The primitive cell of $ErH_2$ crystal were displayed. Green spheres represented Er, while gray spheres represented H. (c) and (d) The DOS at 14.5 and 15 GPa respectively. (e) The variation of spin DOS at $E_F$ and the spontaneous magnetic susceptibility ($\chi_s$) along with the applied pressures.

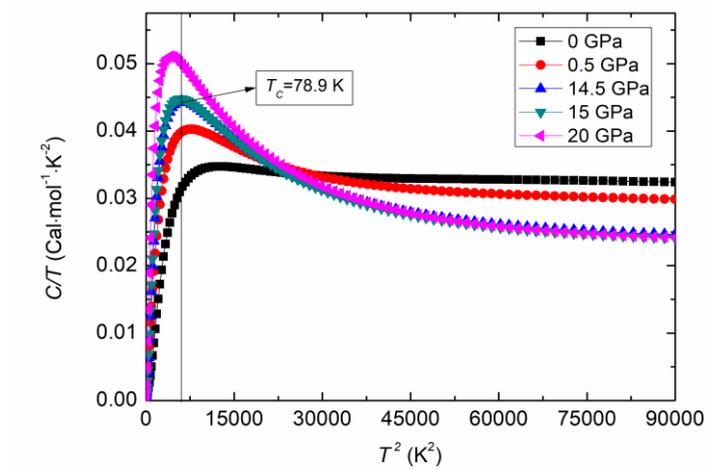

**Figure 5.** $T^2$ dependence of the specific heat in the form of $C/T$ under ambient pressures for ErH$_2$.

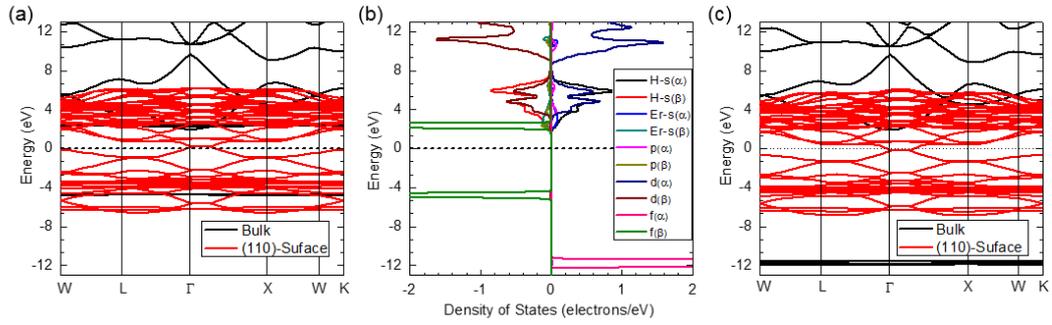

**Figure 6.** (a) and (c) The band structures of this bulk at 20 GPa with electronic spin down and up respectively. The band structures of (110) surface being built with 1×2×1 supercell of this bulk at 20 GPa were also plotted with red color in (a) and (c). The energy scales of this metallic surface state were from -6 to 6 eV. (b) The DOS of this bulk at 20 GPa. $\alpha$ and $\beta$ denoted spin up and down respectively.

# Supplemental Material for "Predicted high-temperature superconductivity in rare earth hydride ErH$_2$ at moderate pressure"

Yiding Liu,[1,2] Qiang Fan,[3] Jianhui Yang,[1] Lili Wang,[4] Weibin Zhang,[5] Gang Yao,[6,*]

[1]*College of Mathematics and Physics, Leshan Normal University, Leshan 614004, China.*
[2] *Institute of Atomic and Molecular Physics, Sichuan University, Chengdu 610065, China.*
[3]*School of New Energy Materials and Chemistry, Leshan Normal University, Leshan 614004, China.*
[4] *Institute of Computer Application, China Academy of Engineering Physics, Mianyang 621900, China.*
[5] *College of Physics and Electronics Information, Yunnan Key Laboratory of Optoelectronic Information Technology, Yunnan Normal University, Kunming, 650500, China.*
[6] *Tsung-Dao Lee Institute, Shanghai Jiao Tong University, Shanghai 200240, China.*

[*]yaogang1257@sjtu.edu.cn

The accompanying behaviors which could also to some extent support the affirmation of superconductivity at 15 GPa were presented in this Supplemental Material. The calculated elastic constants $C_{ij}$ under ambient pressures were presented in Table S1. Judged from the mechanical stability criterion,[1] this bulk became mechanical unstable at 14.5 GPa and up to 21.5 GPa except 15 GPa, as shown in Table S1. When the applied pressure was higher than 21.5 GPa, this bulk became lattice dynamical unstable, i.e. imaginary frequency of phonon dispersions (not show) which was also the reason why the upper limit of the applied pressure in this work was 21.5 GPa. The mechanical stability criterion for cubic crystal could be expressed as:[1] $\tilde{C}_{44} > 0$, $\tilde{C}_{11} > |\tilde{C}_{12}|$, $\tilde{C}_{11} + 2\tilde{C}_{12} > 0$, where $\tilde{C}_{\alpha\alpha} = C_{\alpha\alpha} - P$ ($\alpha = 1, 4$) and $\tilde{C}_{12} = C_{12} + P$, where $P$ is the applied pressure. Meanwhile, superconductivity could also be reflected by an anomalous elastic softening over a temperature range for PuCoGa$_5$ which was a heavy fermion superconductor with $T_c$ = 18.5 K.[2] Then, the analogous hopping of mechanical stability between 14.5 and 15 GPa of ErH$_2$ (Table S1) could also support the deduction of superconductivity at 15 GPa.

By means of Birch–Murnaghan Equation of State (EOS) fitting, the normalized molar volume ($V/V_0$) of ErH$_2$ at 300 K as a function of pressures (Fig. S1) was obtained by GIBBS thermodynamics scheme[3] using single point energy of unit cell ($E$) versus its volume ($V$) within DFT, i.e. *E-V* curve which could be produced in this work (Fig. S2). The corresponding $V/V_0$ versus pressures relation calculated directly with DFT of this work was also plotted in Fig. S1. There was a tiny and clear deviation at this pressure away from the monotonous trend of the fitted curve in this work. Both the *E-V* data and the $V/V_0$ versus pressures relation of DFT had the same parameter settings and precision with the geometry optimizations of its unit cell (see **2. Methods** of the main text). The comparison with experimental data for $ErH_{2.091}$ and $ErH_{1.95}$ measured at room temperature using diamond anvil cell techniques (DAC)[4] was also shown in Fig. S1.

The hypothesis of superconductivity at 20 GPa could also be supported by the tiny and clear deviation of the $V/V_0$ versus pressures relation calculated directly with DFT at this pressure away from the monotonous trend of the curve fitted with *E-V* data of this work (see Fig. S1).

## Supplemental Figures

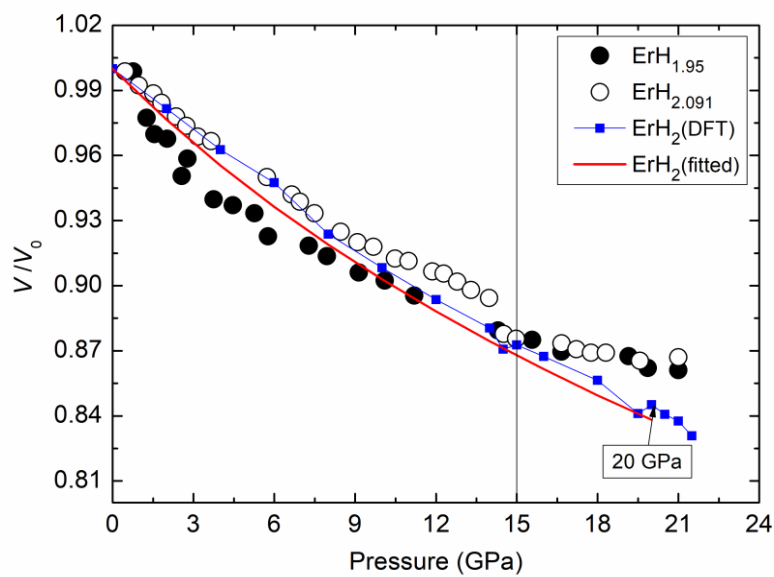

**Fig. S1** The fitted normalized molar volumes ($V/V_0$) as a function of pressures for ErH$_2$ at 300 K (red solid line) using GIBBS thermodynamics scheme[3] with *E-V* data of DFT in this work. The corresponding relation calculated directly with DFT of this work was also plotted (blue solid line with square). The DAC data from Ref. 4 measured at room temperature for ErH$_{1.95}$ (solid circle) and ErH$_{2.091}$ (hollow circle) were also shown with denotion in the inset.

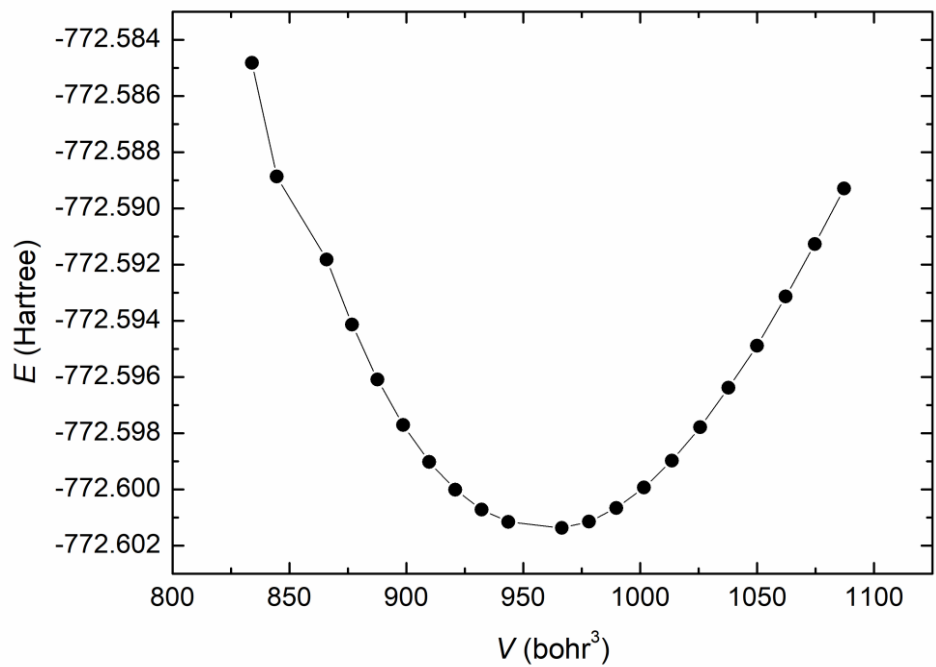

**Fig. S2** The single point energy for unit cell ($E$) of ErH$_2$ versus its volume ($V$) within DFT.

## Supplemental Table

**Table S1.** The calculated elastic constants $C_{ij}$ of ErH$_2$ under ambient pressures (in GPa) and the satisfaction with mechanical stability criteria (Y and N denoted the satisfaction and the unsatisfaction respectively).

| P    | $C_{11}$ | $C_{12}$ | $C_{44}$ | Y/N |
|------|----------|----------|----------|-----|
| 0    | 121.637  | 51.209   | 64.972   | Y   |
| 6    | 158.495  | 59.177   | 54.465   | Y   |
| 12   | 133.738  | 108.377  | 113.144  | Y   |
| 14   | 145.060  | 101.757  | 253.158  | Y   |
| 14.5 | 127.776  | 161.299  | 192.400  | N   |
| 15   | 164.900  | 103.873  | 119.387  | Y   |
| 16   | 148.068  | 126.347  | 76.245   | N   |
| 18   | 176.988  | 105.421  | -50.332  | N   |
| 20   | 129.393  | 192.898  | -24.491  | N   |